\documentclass[aps,prd,reprint,groupedaddress,showkeys,longbibliography,preprintnumbers,nofootinbib]{revtex4-2}
\usepackage[utf8]{inputenc}
\usepackage[english]{babel}
\usepackage[
  bookmarks=false,
  pdfpagelabels=false
]{hyperref}
\usepackage{url}
\usepackage{color}
\usepackage{amsmath}
\usepackage{amsfonts}
\usepackage{amssymb}
\usepackage{graphicx}
\usepackage[left=2cm,right=2cm,top=2cm,bottom=2cm]{geometry}

\newcommand{\dx}{\mathrm{d}}
\newcommand{\eps}{\varepsilon}
\newcommand{\dilog}{\mathrm{Li}_2}
\newcommand{\trilog}{\mathrm{Li}_3}
\newcommand{\qb}{\bar{q}}
\newcommand{\qp}{q^\prime}
\newcommand{\qbp}{\qb^\prime}
\newcommand{\ghy}{\!\,_2\mathrm{F}_1}
\newcommand{\iu}{\mathrm{i}}
\newcommand{\iz}{\mathrm{i}0}
\newcommand{\svp}{\mathcal{L}}
\newcommand{\NC}{N_\text{C}}
\newcommand{\change}[1]{{\textcolor{black}{#1}}}
\allowdisplaybreaks
\begin{document}

\preprint{INT-PUB-25-017}

\title{Single-valued representation of unpolarized and polarized semi-inclusive deep inelastic scattering at next-to-next-to-leading order}
\author{Juliane Haug}
\email[]{juliane-clara-celine.haug@uni-tuebingen.de}
\author{Fabian Wunder}
\email[]{fabian.wunder@uni-tuebingen.de}
\affiliation{Institute for Theoretical Physics, University of T\"ubingen\\
Auf der Morgenstelle 14, 72076 T\"ubingen, Germany}
\affiliation{Institute for Nuclear Theory (INT), University of Washington, Seattle, Washington 98195-1550, USA}

\date{October 24, 2025}

\begin{abstract}
We revisit the recently published analytic results for unpolarized and polarized semi-inclusive deep inelastic scattering (SIDIS) at next-to-next-to-leading order (NNLO) in QCD.
These expressions for the hard scattering coefficients contain case distinctions in the kinematic $(x,z)$-plane, splitting the analytic result into four regions.
By re-expressing the coefficient functions in terms of single-valued polylogarithms, we remove these case distinctions and can present a unified result valid across the entire kinematic range of SIDIS.
This reduces the length of the overall expressions by 30\% to 60\%.
\end{abstract}


\maketitle
\section{Introduction}
Semi-inclusive deep inelastic scattering (SIDIS) is one of the cornerstones of the physics program at the upcoming Electron-Ion Collider (EIC) \cite{Accardi:2012,EIC:2022}.
By probing hadrons created in high-energy scattering of (polarized) electrons off (polarized) protons, SIDIS allows access to (polarized) parton distributions (PDFs) and fragmentation functions (FFs) \cite{Aschenauer:2019}.
In the context of PDF and FF determination, SIDIS is of particular importance to achieve flavor separation.
At present, SIDIS data exist from several experiments \cite{EuropeanMuon:1991sne,ZEUS:1995,H1:1996muf,HERMES:2012uyd,COMPASS:2016xvm}.

The theoretical description of SIDIS is based on the factorization of the process into long-distance non-perturbative hadron dynamics captured by PDFs and FFs and short-distance hard scattering coefficient functions, which are accessible in perturbative quantum chromodynamics (pQCD).
To achieve high precision in the PDF and FF extraction, relevant also for new physics searches, especially in the context of the EIC \cite{Hammou:2024xuj}, a highly accurate calculation of the perturbative part is paramount.
Hence, it has been an important achievement that SIDIS coefficient functions have recently become available to full next-to-next-to-leading order (NNLO) for both the unpolarized and polarized case \cite{Goyal:2023unpol,Bonino:2024unpol,Bonino:2024pol,Goyal:2024pol,Goyal:2024emo}.
This allows for the inclusion of SIDIS data in future fully NNLO fits of polarized parton distributions.
So far, only approximate NNLO based on threshold resummation \cite{Abele:2021,Abele:2022} could be used, necessitating cuts on the data included in the MAPPDFpol1.0 \cite{Bertone:2024taw} and BDSSV24 \cite{Borsa:2024mss} fits while NNPDFpol2.0 \cite{Cruz-Martinez:2025ahf} did not include SIDIS data.

\begin{figure}
		\includegraphics[width=0.75\columnwidth]{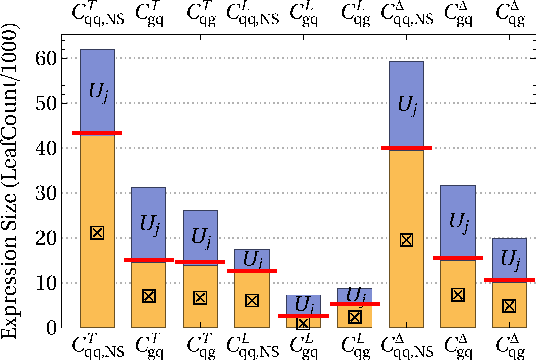}
		\vspace{-0.1cm}
		\caption{Comparison of expression sizes before (full bars) and after (horizontal red lines) removing case distinctions. $U_j$ indicates all parts of the expressions from \cite{Bonino:2024unpol,Bonino:2024pol} involving case distinctions, $\boxtimes$ the rest (see eq.~\eqref{eq: SIDIS regions}).
		Expressions are consistently sorted, \texttt{LeafCount} (number of indivisible subexpressions in \texttt{Mathematica}) serves as proxy to measure expression length.}
		\label{fig: Length Comparison}
		\vspace{-0.4cm}
	\end{figure}

In this {article}, we revisit the analytic NNLO results for SIDIS. These contain case distinctions in the kinematic ($x$,$z$)-plane, splitting the result into four regions.
By re-expressing the coefficient functions in terms of a particular version of single-valued polylogarithms (SVPs), we remove these case distinctions and present a result valid across the entire kinematic range of SIDIS.
As depicted in figure \ref{fig: Length Comparison}, this substantially reduces the length of the overall coefficient functions by 30\% to 60\%, nearly reducing it to the length of the part of the original expressions which does not involve case distinctions.
The benefit of this is twofold.
Phenomenologically, in the context of PDF fits it is of interest to have compact expressions that can be calculated quickly since they need to be evaluated millions of times.
Closely related, due to the absence of case distinctions, our results may help establishing an analytic Mellin transform of the cross section, highly desirable for PDF extraction.
More formally, it is intriguing to see SVPs appearing for the first time in the simplification of observable level quantities in QCD beyond the Regge limit \cite{DelDuca:2013lma}, an idea that might generalize to other processes and higher-loop calculations.
So far, SVPs have seen use on amplitudes \cite{Dixon:2012,Chavez:2012,Dixon:2019,DelDuca:2022,Abreu:2024} and conformal theories \cite{Yan:2022}.

The remainder of this \change{article} is structured as follows.
In section \ref{sec: SIDIS fundamentals}, we recall the fundamentals of the SIDIS process and introduce essential notation.
Readers familiar with \cite{Bonino:2024unpol,Bonino:2024pol} might skip directly to section \ref{sec: Case distinctions}, where the case distinctions present in the NNLO coefficient functions are discussed.
We proceed by removing these through introducing SVPs in section \ref{sec: Single valued polylogarithms for SIDIS}, before concluding in section \ref{sec: Conclusion}.
\change{Appendix \ref{app:Numerical evaluation} discusses to what extent the more compact expressions lead to faster numerical evaluation.
Based on these findings, we introduce the \texttt{C++} special functions library \texttt{BEAVER} in Appendix \ref{app:Beaver}.}
\change{Appendix \ref{app:Mellin} briefly explores the benefits of SVPs in the analytic calculation of double Mellin moments.}
The results and a numerical implementation of SVPs are provided as \texttt{Mathematica} readable ancillary files.
\change{The functions of the \texttt{BEAVER} library are also included with the submission.}

\section{SIDIS kinematics, factorization, and coefficient functions}
\label{sec: SIDIS fundamentals}
	\begin{figure}
	\centering
	\includegraphics[width=0.74\columnwidth]{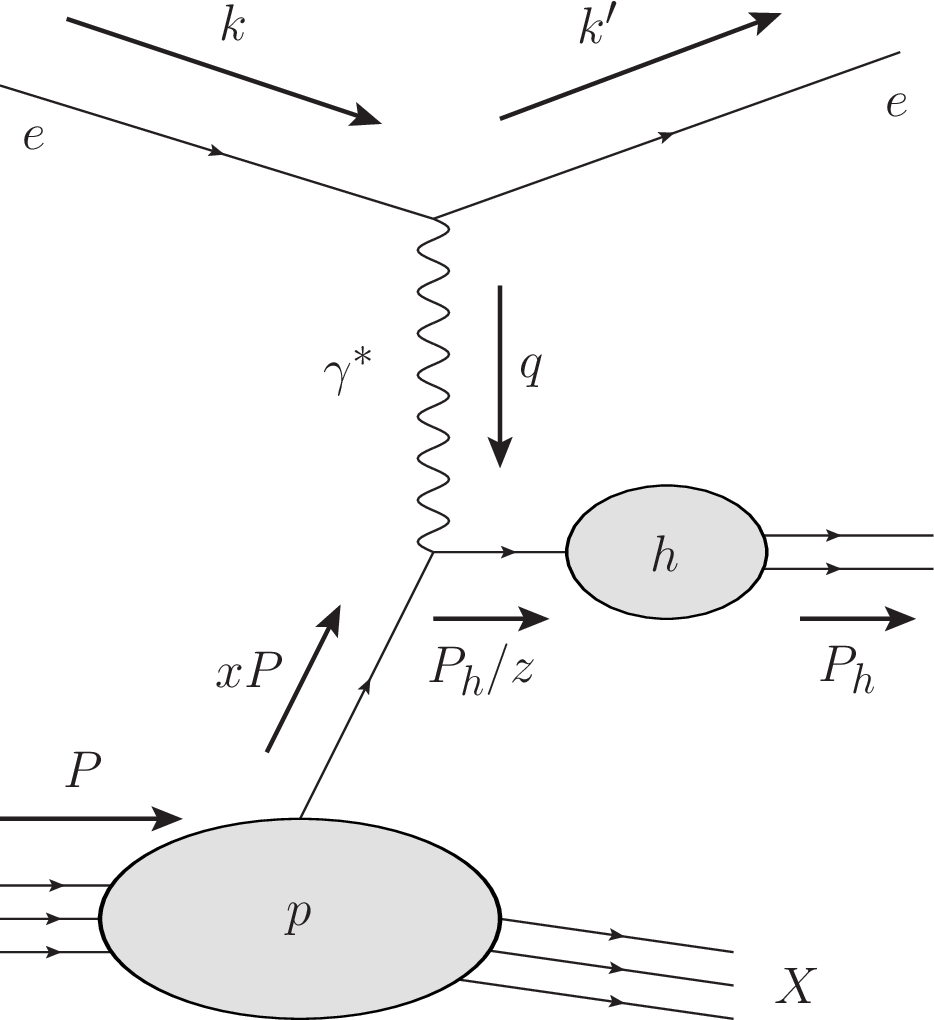}
	\caption{Sketch of the SIDIS process $e(k)\,p(P)\rightarrow e(k^\prime)\,h(P_h)+X$ in the parton model. Here, $e$ denotes the scattered electron,  $\gamma^\ast$ the intermediate virtual photon, $p$ the incoming proton, $h$ the identified hadron, and $X$ the proton remnant (including all unidentified particles in the final state). In SIDIS the electron momentum and the identified hadron's longitudinal momentum are measured.
	Radiative corrections are not included in the picture.}
	\label{fig: SIDIS_Kinematics}
	\end{figure}
The kinematics of the process $e(k)\,p(P)\rightarrow e(k^\prime)\,h(P_h)+X$, as sketched in Fig.~\ref{fig: SIDIS_Kinematics}, is determined by the invariant mass of the intermediate photon $Q^2=-q^2$, where $q=k-k^\prime$ is fixed by the observed electron, the energy transfer of the electron to the proton $y=(P\cdot q)/(P\cdot k)$ and the hadronic variables
	\begin{align}
		x_\change{B}=\frac{Q^2}{2P\cdot q},\quad
		z_\change{h}=\frac{P\cdot P_h}{P\cdot q}\,.
	\end{align}
The Bjorken variable $x_B$, familiar from inclusive DIS, characterizes the longitudinal momentum fraction of the parton within the incoming proton, while the fragmentation variable $z_h$ describes the longitudinal momentum fraction of the outgoing parton carried by the identified hadron. 
Note that SIDIS as discussed here is inclusive in the observed hadron's transverse momentum.
The triple differential cross section for the unpolarized case reads
	\begin{align}
		\frac{\dx^3\sigma^h}{\dx x_B \, \dx y \, \dx z_h}=\frac{4\pi \alpha^2}{Q^2}&
		\left[\frac{1+(1-y)^2}{2y}\,\mathcal{F}_\text{T}^h(x_B,\change{z_h},Q^2)\right.
		\nonumber\\
		&\left.
		+\, \frac{1-y}{y}\,
		\mathcal{F}_\text{L}^h(x_B,\change{z_h},Q^2)
		\right],
	\end{align}
where $\alpha$ is the electromagnetic fine-structure constant and $\mathcal{F}_{\text{T}/\text{L}}^h$ are transverse and longitudinal structure functions, respectively.
The structure functions can be collinearly factorized, up to higher-twist corrections, in the form
	\begin{align}
		&\mathcal{F}^h_{\text{T}/\text{L}}(x_B,z_h,Q^2)=\sum_{a,b}\int_{x_B}^1\frac{\dx 	x}{x}\int_{z_h}^1\frac{\dx z}{z}\,f_{a/p}\!\left(\frac{x_B}{x},\mu_F^2\right)
		\nonumber\\
		&\qquad\times 		D_{h/b}\!\left(\frac{z_h}{z},\mu_A^2\right)\mathcal{C}_{ba}^{\text{T}/\text{L}}\!\left(x,z,\mu_R^2,\mu_F^2,\mu_A^2\right) ,
		\label{eq: Factorization unpol}
	\end{align}
where the sum runs over all partons $a,b= q,\bar{q}, g$.
Here, $f_{a/p}$ denotes the PDF for finding parton $a$ in the proton, $D_{h/b}$ denotes the FF for the fragmentation of parton $b$ into hadron $h$; $\mu_\text{R}^2$, $\mu_\text{F}^2$, and $\mu_\text{A}^2$ are the renormalization, factorization, and fragmentation scales, respectively.

For longitudinally polarized incoming electrons and protons producing an unpolarized hadron in the final state, the analogous differential cross section
	\begin{align}
		\frac{\dx^3\Delta\sigma^h}{\dx x_B \, \dx y \, \dx z_h}=\frac{4\pi 	\alpha^2}{Q^2}(2-y)\,g_1\!\left(x_B,z_h,Q^2\right)
	\end{align}
is parametrized by a single polarized structure function $g_1$.
The factorization analogous to eq.~\eqref{eq: Factorization unpol} in the polarized case reads
	\begin{align}
		&2g_1\!\left(x_B,z_h,Q^2\right) =\sum_{a,b}\int_{x_B}^1\frac{\dx x}{x}\int_{z_h}^1\frac{\dx z}{z}\,\Delta f_{a/p}\!\left(\frac{x_B}{x},\mu_F^2\right)
		\nonumber\\
		&\qquad\times D_{h/b}\!\left(\frac{z_h}{z},\mu_A^2\right)\mathcal{C}_{ba}^{\Delta}\!\left(x,z,\mu_R^2,\mu_F^2,\mu_A^2\right)\,,
		\label{eq: Factorization pol}
	\end{align}
where $\Delta f_{a/p}$ denotes the polarized PDF.
The partonic coefficient functions in both unpolarized and polarized cases only depend on the hard interaction with the virtual photon and can be calculated in perturbative QCD. 
The expansion up to NNLO in the strong coupling $\alpha_s$ reads
\begin{align}
\mathcal{C}_{ba}^{i}=\mathcal{C}_{ba}^{i,(0)}\!+\!\frac{\alpha_s\!\left(\mu_R^2\right)}{2\pi}\,\mathcal{C}_{ba}^{i,(1)}
\!+\!\left(\frac{\alpha_s\!\left(\mu_R^2\right)\!}{2\pi}\right)^{\!\!2}\!\mathcal{C}_{ba}^{i,(2)}\!+\!\mathcal{O}\!\left(\alpha_s^3\right) ,
\end{align} 
where $i=\text{T},\text{L},\Delta$.
The coefficients up to NLO have been known for some time \change{\cite{Altarelli:1979kv, Baier:1979sp, Furmanski:1981, deFlorian:1997}}.
Following the notation of \cite{Anderle:2016,Bonino:2024unpol,Bonino:2024pol}, the seven partonic channels at NNLO are
\begin{align}
C^{i,(2)}_{qq}&=C^{i,(2)}_{\qb\qb}=e_q^2 C^{i,\mathrm{NS}}_{qq}+\left( \sum_j e^2_{q_j}\right)C^{i,\mathrm{PS}}_{qq} \, , \nonumber \\
C^{i,(2)}_{\qb q}&=C^{i,(2)}_{q \qb}=e_q^2C^{i}_{\qb q} \, , \nonumber \\
C^{i,(2)}_{\qp q}&=C^{i,(2)}_{\qbp \qb}=e_q^2 C^{i,1}_{\qp q}+e_{\qp}^2 C^{i,2}_{\qp q}+e_q e_{\qp} C^{i,3}_{\qp q} \, , \nonumber \\
C^{i,(2)}_{\qbp q}&=C^{i,(2)}_{\qp \qb}=e_q^2 C^{i,1}_{\qp q}+e_{\qp}^2 C^{i,2}_{\qp q}-e_q e_{\qp} C^{i,3}_{\qp q} \, , \nonumber \\
C^{i,(2)}_{gq}&=C^{i,(2)}_{g \qb}=e_q^2 C^{i}_{gq} \, , \nonumber \\
C^{i,(2)}_{qg}&=C^{i,(2)}_{\qb g}=e_q^2 C^{i}_{qg} \, , \nonumber \\
C^{i,(2)}_{gg}&=\left( \sum_j e_{q_j}^2 \right) C^{i}_{gg} \,  ,
\label{eq: CFNNLOlist}
\end{align}
where $q$ and $\qp$ denote (anti-)quarks of different flavor, NS is the non-singlet and PS the pure-singlet contribution.

\section{Case distinctions in NNLO SIDIS cross section}
\label{sec: Case distinctions}
The coefficients of eq.~\eqref{eq: CFNNLOlist} have been recently calculated independently by two groups in \cite{Goyal:2023unpol,Bonino:2024unpol,Bonino:2024pol,Goyal:2024pol,Goyal:2024emo}.
The results of both are in (numerical) agreement and available as supplemental material to the aforementioned publications.
For both the unpolarized and polarized structure functions, the expressions are of considerable length.
Overall the \change{published} results of \cite{Bonino:2024unpol,Bonino:2024pol} are more compact compared to \cite{Goyal:2024emo}, most importantly with regard to a minimal number of polylogarithms.
Hence, we choose the former as baseline for our simplified results; however, we explicitly checked that the presented method works analogously for the latter.

\begin{figure}
		\includegraphics[width=0.7\columnwidth]{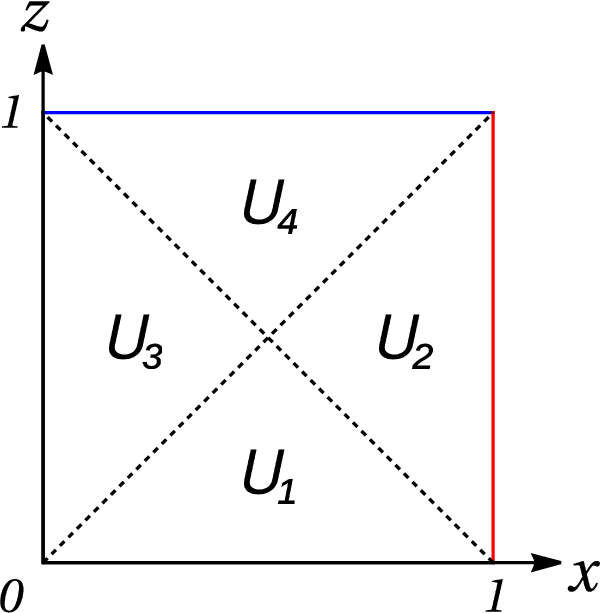}
		\caption{The four regions of SIDIS for which there are case distinctions in the NNLO coefficient functions as presented in \cite{Goyal:2023unpol,Bonino:2024unpol,Bonino:2024pol,Goyal:2024pol,Goyal:2024emo}.
		The red and blue boundaries depict the kinematic endpoints for $x$ and $z$ where physical thresholds are present.}
		\label{fig:SIDIS_Regions}
	\end{figure}

Notably, in the channels $q\rightarrow q$ (NS), $q\rightarrow g$, and $g\rightarrow q$, the analytic representation requires case distinctions between four regions in the $(x,z)$-plane.
These regions are depicted in figure \ref{fig:SIDIS_Regions} and follow the notation of \cite{Gehrmann:2022cih}, which was applied in \cite{Bonino:2024unpol,Bonino:2024pol}.
In the language of \cite{Knuth:1992}, they are given by
	\begin{align}
		U_1&=[ x\geq z, x\leq 1-z]\,,\;\;
		U_2=[x\geq z, x\geq 1-z]\,,\nonumber\\
		U_3&=[x\leq z, x\leq 1-z]\,,\;\;
		U_4=[x\leq z, x\geq 1-z]\,.
	\end{align}
Furthermore, the combined regions $\boxtimes=\sum_{j=1}^4 U_j$, $R_1=U_1+U_3$, $\change{R_2}=U_2+U_4$, $T_1=U_3+U_4$, and $T_2=U_1+U_2$ are used to present the coefficient functions.
Hence, the coefficient functions in \cite{Bonino:2024unpol,Bonino:2024pol} have the generic form
	\begin{align}
		C_{ba}^{i,(2)}&=C_{ba}^{i,(2),\boxtimes}+\sum_{\change{j}=1}^4 U_j\,C_{ba}^{i,(2),U_j}
		\nonumber\\
		&+\sum_{\change{j}=1}^2 R_j\,C_{ba}^{i,(2),R_j}+\sum_{j=1}^2 T_j\,C_{ba}^{i,(2),T_j} \,,
		\label{eq: SIDIS regions}
	\end{align}
with different contributions $C_{ba}^{i,(2),r}$ from the different regions $r=\boxtimes,R_j, T_j, U_j$ to the overall coefficient function $C_{ba}^{i,(2)}$.
The aim of this \change{work} is to eliminate the distinction between regions and obtain a single-valued representation across the entire SIDIS range $0\leq x,z\leq 1$.
To achieve this, we start from the origin of the case distinctions.

\begin{figure}
		\includegraphics[width=0.7\columnwidth]{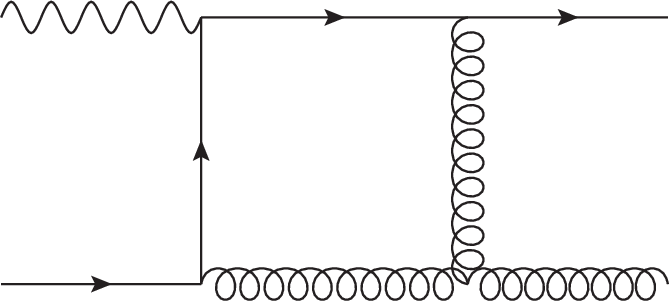}
		\caption{Illustrative one-loop box diagram contributing to the real-virtual correction at NNLO responsible for the spurious branch cuts.}
		\label{fig: box}
	\end{figure}

\begin{figure}
		\includegraphics[width=0.9\columnwidth]{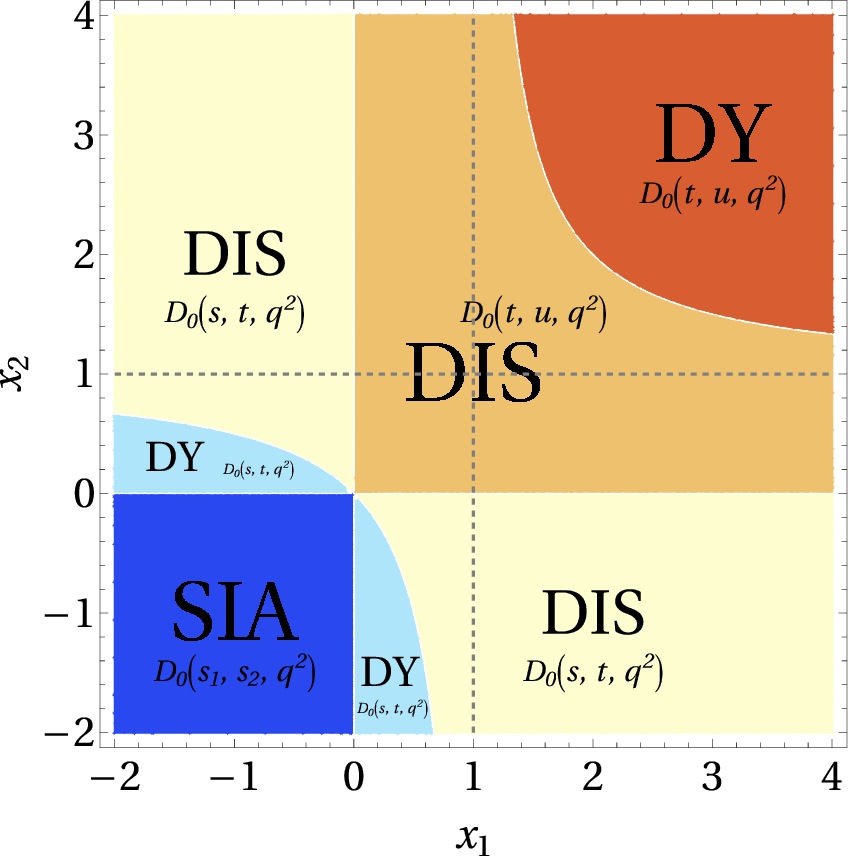}
		\caption{Branch cut structure of the one loop box integral. Each color coded region boundary corresponds to a physical branch cut. $x_{1,2}$ are ratios of Mandelstam variables defined in \cite{Haug:2022}. The dashed gray lines indicate the spurious branch cuts at $x_i=1$ induced by the hypergeometric functions in the representation of eq.~\eqref{eq: Scalar box integral} running right through the DIS regions.}
		\label{fig: Box integral regions}
	\end{figure}

At NNLO, there are three classes of QCD corrections, virtual-virtual, real-virtual, and real-real.
The case distinctions are only present in the channels that receive real-virtual corrections, i.e., those already present at NLO.
The reason behind this is that the case distinctions originate solely from  one-loop box diagrams such as the one depicted in figure \ref{fig: box}.
The relevant analytical structure comes from the scalar box integral $D_0$, which in $d=4-2\eps$ dimensions has the general form \cite{Fabricius:1979, Matsuura:1988, Bern:1993, Zijlstra:1992, Duplancic:2000, Lyubovitskij:2021, Haug:2022, Haug:2023}, valid not only for SIDIS kinematics but also for Drell-Yan (DY) and single-inclusive annihilation (SIA),
	\begin{align}
		D_0(s_1,s_2,q^2)\sim& \left(\!\frac{\mu^2}{-s_2}\!\right)^{\!\eps}\!\mathrm{F}_\eps\!\left(\frac{-s_3}{s_1}\right)
		\!+\!\left(\!\frac{\mu^2}{-s_1}\!\right)^{\!\eps}\!\mathrm{F}_\eps\!\left(\frac{-s_3}{s_2}\right)
		\nonumber\\&
		-\left(\frac{\mu^2}{-q^2}\right)^{\!\eps}\!\mathrm{F}_\eps\!\left(\frac{-s_3 q^2}{s_1 s_2}\right).
		\label{eq: Scalar box integral}
	\end{align}
Here $\mathrm{F}_\eps(z)=\ghy(1,-\eps,1-\eps;z)$ abbreviates a particular Gauss hypergeometric function, the $s_i$ are Mandelstam variables, $q^2$ is the invariant mass of the external off-shell particle, and $\mu^2$ the renormalization scale.
The prefactors $(\dots)^\eps$ in eq.~\eqref{eq: Scalar box integral} develop branch cuts when the respective Mandelstam variables or $q^2$ become positive.
A change in sign of Mandelstam variables is associated with an interchange of incoming and outgoing particles, i.e., physically different kinematics shown in figure \ref{fig: Box integral regions}.
Therefore, those branch cuts are physical and the associated factors of $\iu \pi$ may contribute in the absolute square of matrix elements as factors of $\pi^2$.
However, the hypergeometric functions $\mathrm{F}_\eps$ in eq.~\eqref{eq: Scalar box integral} develop branch cuts when their arguments become \change{larger than} unity, for SIDIS these arguments are
	\begin{align}
		\frac{1-x}{z}\,,\,\,
		\frac{1-x}{1-z}\,,\,\,
		\frac{(1-x)x}{(1-z)z}\,,\,\,
		\frac{x z}{(1-x)(1-z)}\,,\,\,
		\frac{(1-z)x}{(1-x)z}\,,
		\label{eq: SIDIS arguments}
	\end{align}
and also the respective inverses of the first three.
They become unity at $x=z$ or $x=1-z$, giving rise to spurious branch cuts within the SIDIS region that always cancel in the sum between the three $\mathrm{F}_\eps$ in eq.~\eqref{eq: Scalar box integral}.
In \cite{Bonino:2024unpol,Bonino:2024pol} and \cite{Goyal:2023unpol,Goyal:2024pol,Goyal:2024emo} those were treated in a similar way by transforming the arguments of the hypergeometric functions away from the region where the arguments become larger than $1$ using hypergeometric transformations.
This necessitates case distinctions based on the regions of figure \ref{fig:SIDIS_Regions} that remain in the final analytic result.
The exact procedure employed by the different groups is detailed in 
section 5 of \cite{Gehrmann:2022cih} and section \change{III.}C of \cite{Goyal:2024emo}, respectively.
\change{We note that while the published results for the coefficient functions of the latter group contain case distinctions, in \cite{Goyal:2024emo} they presented master integrals for the real-virtual corrections in terms of hypergeometric $_2F_1$ functions, which are analytic in the whole physical region.
We will comment on the relation to our results at the end of Section \ref{sec: Single valued polylogarithms for SIDIS}.}

In \cite{Haug:2022,Haug:2023}, we have discussed how to treat the spurious branch cuts in eq.~\eqref{eq: Scalar box integral} without introducing case distinctions by using a particular version of SVPs.
In the next section we will recall this approach and apply it to the NNLO SIDIS coefficient functions

\section{Expressing SIDIS coefficient functions with single-valued polylogarithms}
\label{sec: Single valued polylogarithms for SIDIS}
\change{Carefully tracking both the causal imaginary part $+\iz$ from the propagators as well as the spurious branch cuts introduced by expressing $D_0$ in terms of $\mathrm{F}_\eps$, we showed in \cite{Haug:2022} that the box integral as given in eq.~\eqref{eq: Scalar box integral} can be written as
	\begin{align}
		D_0&(s_1,s_2,q^2)\sim 
		\left(\frac{\mu^2}{-s_2-\iu 0}\right)^{\!\eps} \, \left|\frac{s_3}{s_1}\right|^\eps \, \mathfrak{F}_{\eps}\!\left(\frac{-s_3}{s_1}\right)
		 \nonumber \\
		& + \left(\frac{\mu^2}{-s_1-\iu0}\right)^{\!\eps} \, \left|\frac{s_3}{s_2}\right|^\eps \, \mathfrak{F}_{\eps}\!\left(\frac{-s_3}{s_2}\right)
		\nonumber\\
		& - \left(\frac{\mu^2}{-q^2-\iu 0}\right)^{\!\eps} \, \left|\frac{s_3 q^2}{s_1
		s_2}\right|^\eps \, \mathfrak{F}_{\eps}\!\left(\frac{-s_3 q^2}{s_1s_2}\right).
	\end{align}
The physical branch cuts are made explicit by the $\iz$ prescription, while the cancellation of spurious branch cuts in eq.~\eqref{eq: Scalar box integral} is achieved by replacing $F_\eps(x)\rightarrow |x|^\eps \mathfrak{F}_\eps(x)$, where each $\mathfrak{F}_\eps(x)$ is an individually single-valued version of the hypergeometric function.
By contrast, in eq.~\eqref{eq: Scalar box integral} the spurious branch cuts only cancel in the sum of hypergeometric functions.}
In particular, \change{$\mathfrak{F}_\eps(x)$} is real valued also for $x>1$ and given by
	\begin{align}
		\mathfrak{F}_\eps(x)\! = \!1 - \ln\!\left|\frac{x}{x-1}\right| \sum_{n=1}^\infty\frac{\eps^n}{n!}\ln^{n-1}\!\left|\frac{1}{x}\right|\! - \!\sum_{n=2}^\infty\eps^n\svp_n(x),
		\label{eq: eps expansion Ffrak}
	\end{align}
where $\svp_n(x)$ is a specific version of SVP \cite{Lewin1991,Haug:2022}
	\begin{align}
		\svp_n(x) = \sum_{k=0}^{n-1}\frac{\ln^k\!\left|\frac{1}{x}\right|}{k!}\, \mathrm{Li}_{n-k}(x)+\frac{\ln^{n-1}\!\left|\frac{1}{x}\right|}{n!}\ln|1-x|\,.
		\label{eq: SVP definition}
	\end{align}
\change{Note that for $x>1$, the polylogarithms are taken on their principal branch (which is inherited from the principal branch of the logarithm), i.e., $\mathrm{Li}_n(x>1)\equiv \mathrm{Li}_n(x-\iz)$.}

SVPs have a rich pedigree in pure mathematics and mathematical physics \cite{Kummer1840,Rogers1907,Ramakrishnan1986,Wojtkowiak1989,
Lewin1991,Zagier1991,Brown:2004,Bloch2011,Schnetz:2013,Brown:2015,Zhao2016} and have become a field of active research in the context of not only amplitude calculations
\cite{Dixon:2012,Chavez:2012,Broedel:2016,Duhr:2019,Dixon:2019,Charlton:2021,DelDuca:2022,Frost:2023,Abreu:2024} but also the Balitsky-Fadin-Kuraev-Lipatov (BFKL) equation \cite{DelDuca:2013lma}, soft-anomalous dimensions \cite{Almelid:2015}, thermal one-point functions \cite{Petkou:2021}, and energy correlators \cite{Yan:2022}.
In the particular situation at hand, the appearance of single-valued functions is not coincidental but a consequence of the cancellation of spurious branch cuts built into the physical box integral.

Besides being single-valued and continuous on $\mathbb{R}$ by construction, the SVPs of eq.~\eqref{eq: SVP definition} have the remarkable property of satisfying clean versions of polylogarithmic functional equations, i.e., free of product terms.
Specifically, the inversion relation takes the form
	\begin{align}
		\svp_n(x)+(-1)^n\svp_\change{n}\!\left(\frac{1}{x}\right)=\svp_n(\mathrm{sgn}(x)\,\infty),
		\label{eq: Inversion relation}
	\end{align}
with constants $\svp_{2n+1}(\pm\infty)=0$, $\svp_{2n}(\infty)=2\zeta_{2n}$ and $\svp_{2n}(-\infty)=2(2^{1-2n}-1)\zeta_{2n}$.
Moreover, these SVPs are bounded on $\mathbb{R}$, which is welcome for numerical evaluation as well as for the study of kinematic limits.
More details on the SVPs of eq.~\eqref{eq: SVP definition} can be found in \cite{Haug:2022}.
While we are led to single-valued versions of classical polylogarithms in this particular case because they appear in the $\eps$-expansion of the one-loop box, demanding cancellation of spurious branch cuts for multi-loop generalizations is expected to naturally lead to single-valued versions of generalized polylogarithms \cite{Schnetz:2013,Charlton:2021}.

In the NNLO SIDIS coefficient functions only polylogarithms of weight $2$ appear together with region distinctions.
Hence, we only need to express the dilogarithm in terms of its single-valued counterpart,
	\begin{align}
		\!\!\!\mathrm{Li}_2(x)\!\rightarrow\!\svp_2(x)-\ln|x|\ln(1-x)+\frac{\ln|x|}{2}\ln|1-x|\,,\!\!
	\end{align}
in all regions and introduce absolute values for all logarithms that develop branch cuts outside their original region.
Using the inversion relation eq.~\eqref{eq: Inversion relation} on certain SVPs, the expression in all regions can be mapped onto a universal form allowing us to write the coefficient functions without case distinctions.
Since the former four regions are now expressed by a single term, the size of the part involving case distinctions shrinks by approximately $75$\%, as one would expect for an expression of similar complexity as the individual contributions to the four regions $U_i$.
Finally, sorting this contribution into the structure of the remaining part of the coefficient without case distinctions $C_{ba}^{i,(2),\boxtimes}$ shrinks the overall coefficient function $C_{ba}^{i,(2)}$ nearly to the size of $C_{ba}^{i,(2),\boxtimes}$ (compare the horizontal red lines to the yellow bars in figure \ref{fig: Length Comparison}), resulting in overall size reductions ranging from 30\% to 60\%.
To consistently and meaningfully compare the sizes of expressions, all coefficients have been sorted in a way closely matching \cite{Bonino:2024unpol,Bonino:2024pol} with slight adaptations to accommodate importing the expressions from \texttt{FORM} to \texttt{Mathematica}.

\change{Before concluding, we want to briefly comment on the relation between our single-valued representation of the result and the master integrals presented in \cite{Goyal:2024emo}, which are case distinction free without using single-valued polylogarithms, and also permit for deriving unified representations of the coefficient functions.
To understand the connection, we observe that the dilogarithm arguments causing issues at the spurious branch cuts, listed in eq.\,\eqref{eq: SIDIS arguments}, are all non-negative.
These arguments, which are those appearing in single-valued dilogarithms in our result, can be mapped with the functional equation $x\rightarrow 1-x$ \cite{Haug:2022} to the interval $-\infty<x<1$.
Subsequently expressing the single-valued dilogarithms as regular dilogarithms, the arguments of the latter are away form the branch cut.
Hence, following these steps, our result and a result using the master integral of \cite{Goyal:2024emo} in a way that avoids spurious branch-cuts, can be mapped onto each other.
For higher orders in the $\eps$-expansion of eq.\,\eqref{eq: eps expansion Ffrak}, which would contribute in a N$^3$LO SIDIS calculation, single-valued trilogarithms appear.
For those, the analogous $x\rightarrow 1-x$ mapping leaves the space of ordinary (single-valued) polylogarithms.
This observation is consistent with eq.\,(49) of \cite{Goyal:2024emo}, where harmonic polylogarithms appear in higher orders of the $\eps$-expansion.}

\section{Conclusion and Outlook}
\label{sec: Conclusion}
In this {article}, we presented a significant compactification of the existing NNLO results for SIDIS.
Leveraging the understanding of how the underlying branch cut structure of the loop integrals introduced spurious case distinctions in the results, we re-expressed the affected parts of the coefficient functions in terms of SVPs, the first use of the latter for observable level quantities in QCD beyond the Regge limit.
This provides a representation both phenomenologically relevant in the context of (polarized) PDF extraction and theoretically interesting with a view to generalizing the applied method to branch cuts in multi-loop multi-variable observables.
All coefficient functions up to NNLO are given in \texttt{Mathematica} readable ancillary text files together with a numerical implementation of SVPs in part based on \cite{Voigt:2022xnc}.

\change{As we discuss in Appendix \ref{app:Numerical evaluation} in more detail, for immediate phenomenological use the numerical implementation of the SIDIS coefficients as used in practice takes most of its computing budget from the evaluation of special functions.
Hence, to achieve substantial speed-up, fast implementations of logarithms and (single-valued) polylogarithms are essential.
To this end we developed the \texttt{C++} library \texttt{BEAVER} (Better Evaluate A Very Efficient Rational), which is discussed in Appendix \ref{app:Beaver} and can be found 
at \href{https://github.com/fabianwunder/beaver}{\texttt{https://github.com/fabianwunder/beaver}}.
This achieves an order-of-magnitude improvement in the evaluation time of the dilogarithms compared to the use of \texttt{CERNLIB}, which translates to five times speed-up for calling the partonic NNLO SIDIS coefficient functions, and results in an overall speed-up by a factor of around two when used for evaluation of the hadronic SIDIS structure functions or when implemented in the SIDIS Mellin grid calculation of the BDSSV code.}

\change{In future studies, obtaining an analytic double Mellin transform of the SIDIS coefficient functions would be desirable.
Here, the presented single-valued form may be of use, as explored in Appendix \ref{app:Mellin}.}

\begin{acknowledgments}
\textbf{Acknowledgments.}---We are grateful to Thomas Gehrmann and Sven Moch for stimulating discussions on the NNLO coefficient functions, to Ignacio Borsa for sharing insights into PDF fitting, helping to test the \texttt{Mathematica} implementation of SVPs, \change{discussing the \texttt{C++} implementation of the coefficient functions, and performing run-time tests with the BDSSV code,} and to Werner Vogelsang for helpful comments on the manuscript.
Special thanks go to Markus L\"ochner for initiating us into the secrets of rln2 in the supplemental material to \cite{Bonino:2024pol}.
We thank Renee Fatemi, Huey-Wen Lin and Werner Vogelsang for organizing the CFNS-INT Joint Program \textit{Precision QCD with the Electron Ion Collider} and the Institute for Nuclear Theory at the University of Washington for its hospitality and the Department of Energy for partial support during the completion of this work.
Figures \ref{fig: SIDIS_Kinematics} and \ref{fig: box} \change{were} drawn with JaxoDraw \cite{Binosi:2003yf}.
This work was supported by Deutsche Forschungsgemeinschaft (DFG) through the Research Unit FOR 2926 (project 409651613).
\end{acknowledgments}

\appendix
\section{Comment on numerical implementation of coefficient functions}
\label{app:Numerical evaluation}

In figure \ref{fig: Length Comparison}, we compared the analytic expression sizes before and after removing case distinctions with everything else left in place, especially regarding sorting or using functional equations.
Since, besides expressing dilogarithms by their single-valued counterparts, no abbreviations have been introduced, this can be regarded as a ``fair" metric for ``analytic complexity".
In phenomenological practice, e.g. in a PDF fit, when one needs to get out actual numbers from the coefficient functions, the relevant metric, however, is computation time.
How the analytic complexity of the coefficient functions translates to the time it takes to compute them numerically is a highly non-trivial question and will in practice depend first and foremost on the actual code implementation, e.g. in \texttt{FORTRAN} or \texttt{C/C++}.
Here, the benefits of the case distinction free form are a direct reduction of code size, and a cleaner handle on the regions $x=z$, $x=1-z$ that suffer from spurious singularities. These are for example of the form
$
{\log (x/z)}/{(x-z)}
$, but often spread over several terms in the analytic expressions, and hence require careful algebraic organization and numerical treatment.
A similar situation occurs at the kinematic endpoints $x,z=1$ with terms like 
$
{\dilog(1-x)}/{(1-x)}
$\,.

In short, the question how much faster the simplified structure functions evaluate compared with the structure functions published by \cite{Bonino:2024unpol,Bonino:2024pol} or \cite{Goyal:2024emo} is ill-posed.
Hence, we decided it would be more practical to pivot to an actual use case to estimate optimization potential.
For this, we took the code to be used in the next generation of the BDSSV fit as a baseline for improvement.

The BDSSV implementation of the SIDIS NNLO coefficients uses a \texttt{C/C++} code based on the results reported in \cite{Bonino:2024pol}, kindly provided by its authors.
This code includes certain optimizations regarding the handling of the algebra and carefully treats the regions near the spurious singularities discussed above, which makes a direct equivalent replacement by the results of the present paper unfeasible.

Even more importantly, upon benchmarking the code performance, it turned out that the bottlenecks of the present implementation are not algebra or branching, but calls to special functions.
Therefore, the first step to actual performance gains lies in optimizing the latter before any other tweaks become important.
To this end, we developed the small \texttt{C++} library \texttt{BEAVER} for the fast calculation of logarithms, polylogarithms, and other special functions appearing in the SIDIS coefficients. 
Details of \texttt{BEAVER} are presented in Appendix \ref{app:Beaver}.

Replacing all special functions in the SIDIS coefficients by their \texttt{BEAVER} counterparts results in an overall speed-up of about a factor of five, benchmarked across 1,000,000 calls at random $(x,z)$ points of all coefficients.
Inside the BDSSV code (run on an Intel Xeon Bronze 3106, 1.70GHz) -- where the SIDIS coefficients are used in convolution with fragmentation functions to pre-compute a Mellin grid for the subsequent fit of polarized PDFs\footnote{Following the method of the DSS fit described in Section III.C of \cite{deFlorian:2007aj} (see also \cite{Stratmann:2001pb}).} -- preliminary testing showed that introducing \texttt{BEAVER} reduces the computation time per Mellin point from about $320$ seconds down to $170$ seconds, an overall speed-up by a factor of $1.9$.
A  very similar speed-up was observed for the calculation of the hadronic SIDIS structure functions in $(x_B,z_h)$-space. 

\section{Better Evaluate A Very Efficient Rational (BEAVER): Dam fast with (poly-)logs}
\label{app:Beaver}
As explained in Appendix \ref{app:Numerical evaluation}, the main reduction in computation time for the SIDIS coefficients in their currently used form comes from a more efficient calculation of special functions.
For this purpose we developed the small \texttt{C++} library \texttt{BEAVER} -- available 
under \href{https://github.com/fabianwunder/beaver}{\texttt{https://github.com/fabianwunder/beaver}} --, which we hope to be of use also beyond the scope of SIDIS.
The presentation here is intended as a preview, a full release will be the subject of future work.

Compared to standard numerical libraries, there is room for a more lightweight implementation since for typical usage in high energy physics, accuracy down to the unit of least precision (ulp) is not a strict requirement, but rather faithfulness at near machine precision, i.e., relative errors at the $10^{-16}$ level and absence of pathological behavior near branch points.
With the \texttt{vdt} \cite{Piparo:2014tga} library, there is a notable precedent in the field of high energy physics (HEP), which was developed to speed up logarithm and trigonometric function calls in HEP experiment software at the LHC.
Building on the older \texttt{Cephes} \cite{CEPHES:2000} library, \texttt{vdt} uses rational approximations, specifically the Padé approximant -- the rational function of fixed degrees with the most derivatives at a point matching the approximated function.

Similarly, at its core \texttt{BEAVER} uses rational mini-max approximations -- the rational function of fixed degrees that minimizes the maximal relative error over an interval.\footnote{The rational mini-max approximation of a function $f(x)$ of numerator/denominator degree $(d_\text{n}/d_\text{d})$ on the interval $[x_\text{min},x_\text{max}]$ can be calculated with \texttt{Mathematica} using the commands
\texttt{Needs["FunctionApproximations`"]} and subsequent
\texttt{ResourceFunction["MiniMaxApproximation"]}
\texttt{[f[x],\{x,\{xmin,xmax\},dn,dd\}]}.}
In the context of efficient polylogarithm calculation, this idea was put forward in \cite{Voigt:2022xnc, Voigt:2023ztf}.
For \texttt{double} precision, the numerator and denominator polynomials are typically chosen to be of degree around six.
To map to compact regions for the mini-max approximation, suitable functional equations are employed, most prominently $x\rightarrow 1/x$.
Additionally, to ensure regular behavior near zeroes and around branch points, Taylor series are bridge between mini-max regions.

Since the mini-max polylogarithm implementations, such as the di- and trilogarithm implementations of Voigt \cite{Voigt:2022xnc, Voigt:2023ztf}, are limited by the calculation of logarithms inside the function call, a standalone fast logarithm, accurate within a few-ulp, is implemented in \texttt{BEAVER}, which uses a combination of mantissa-exponent decomposition, precomputed tables, and short Taylor series as well as a mini-max rational near $x=1$.
Table \ref{tab: beaver timing} compares evaluation times of the \texttt{BEAVER} functions to the implementations previously used in the SIDIS coefficient functions and also, for the cases of logarithm, arctangent, di- and trilogarithm, to the fast implementations of \cite{Piparo:2014tga} and \cite{Voigt:2022xnc, Voigt:2023ztf}, respectively.

	\begin{table}
		\begin{tabular}{llrr}
			\hline \hline
			Function & Implementation & time/call & \,\,speed-up\\
			\hline
			$\ln(x)$ & \texttt{std::log} & 16.0\,ns & 1 \\
			$\ln(x)$ & \texttt{vdt::fast\_{}log} & 6.7\,ns & 2.3\\
			$\ln(x)$ & \texttt{beaver::log} & 5.0\,ns & $\mathbf{3.2}$ \\
			\noalign{\smallskip}
			$\arctan(x)$ &\texttt{std::atan} & 18.2\,ns & 1\\
			$\arctan(x)$ & \texttt{vdt::fast\_{}atan} & 9.2\,ns & 2.0 \\
			$\arctan(x)$ & \texttt{beaver::atan} & 8.8\,ns & $\mathbf{2.1}$\\
			\noalign{\smallskip}
			$\dilog(x)$ & \texttt{CERNLIB::dilog} & 124.4\,ns & 1 \\
			$\dilog(x)$ & \texttt{voigt::dilog} & 43.6\,ns & 2.9 \\
			$\dilog(x)$ & \texttt{beaver::dilog} & 11.3\,ns & $\mathbf{11.2}$\\
			\noalign{\smallskip}
			$\mathcal{L}_2(x)$ & \texttt{beaver::svdilog} & 14.1\,ns & \dots \\
			\noalign{\smallskip}
			$\trilog(x)$ & \texttt{CERNLIB::trilog} & 132.1\,ns & 1\\
			$\trilog(x)$ & \texttt{voigt::trilog} & 28.2\,ns &  4.7 \\
			$\trilog(x)$ & \texttt{beaver::trilog} & 13.7\,ns & $\mathbf{9.6}$ \\
			\noalign{\smallskip}
			$\mathrm{Ti}_2(x)$ &\texttt{gsl\_{}sf\_{}atanint} & 80.0\,ns & 1\\
			$\mathrm{Ti}_2(x)$ &\texttt{beaver::atanint} & 11.9\,ns & $\mathbf{6.7}$\\
			\hline \hline
			\end{tabular}
		\caption{Result of timing benchmark on 5,000,000 random points in the domain of the respective functions. The speed-up factor in the last column is relative to the implementation used in the current version of the SIDIS coefficients at the top of each block, in bold the speed-up factor of the \texttt{BEAVER} implementation. Existing fast implementations of \texttt{vdt} \cite{Piparo:2014tga}  and \texttt{voigt} \cite{Voigt:2022xnc, Voigt:2023ztf} are given for reference. 
		Benchmark used a standard laptop (ThinkPad T495 with AMD Ryzen 5 PRO 3500U) with a 2.1\,GHz processor (3.7 GHz in boost), compiled with \texttt{g++ -O3 -DNDEBUG -march=native -ffp-contract=fast -fno-trapping-math -fno-math-errno}. 
		Numbers are intended as guidelines, speed-up and especially absolute times may vary according to set-up.}
		\label{tab: beaver timing}
	\end{table}
	
These numbers are intended as broad reference points since there is a considerable dependence on the set-up, on the technical side with regards to the processor and compiler, but also on the range of arguments since computation speed is typically non-uniform on the argument.
Nevertheless, we observe a substantial and stable speed-up compared to the \texttt{std}, \texttt{CERNLIB}, and \texttt{gsl} implementations used in the implementation of the SIDIS coefficients.
Compared to the polylogarithm implementations of Voigt, \texttt{BEAVER} is faster mainly due to the faster logarithm used within the argument transformations.
The faster evaluation of \texttt{beaver::log} compared to \texttt{vdt::fast\_{}log} is achieved by using a 128-bin grid for inputs where the logarithm is not close to its zero.
Regarding accuracy, \texttt{std} and \texttt{gsl} are (nearly) ulp-exact, \texttt{vdt} is very close, \texttt{BEAVER} has single-digit ulp inaccuracies, \texttt{CERNLIB} typically triple-digit.

\section{Towards an analytic calculation of double Mellin moments}\label{app:Mellin}
In this appendix, we briefly showcase how the representation in terms of SVPs can be useful in the analytic calculation of double Mellin moments.
These are defined as
	\begin{align}
		\tilde{\mathcal{C}}(N,M)=\int_0^1\dx x\,x^{N-1} \int_0^1\dx z\,z^{M-1} \mathcal{C}(x,z)\,.
	\end{align}
Since the Mellin transformation turns the convolution of eq.~\eqref{eq: Factorization unpol} and \eqref{eq: Factorization pol} respectively into products of Mellin transforms, Mellin transforming the hadronic structure functions requires the Mellin transformation of the partonic coefficient functions.
These are an invaluable ingredient for PDF extractions, which are often done in Mellin space.

As mentioned in Appendix \ref{app:Numerical evaluation}, the BDSSV code for example uses a numerical Mellin-grid of the convolution of the structure functions with the fragmentation functions.
If an analytic form of the coefficient functions in Mellin space were available, this computationally expensive step could be bypassed.
Calculations in the threshold limit have been performed in Mellin space up to order $\alpha_s^4$ \citep{Abele:2021, Abele:2022,Goyal:2025}.
Transforming the complete non-distributional remainder terms into analytic Mellin-space expressions on the other hand has only been done for the NLO SIDIS coefficients \cite{Stratmann:2001pb} and is an open problem at NNLO.
Here, we observe a stark contrast between the maturity Mellin space methods \cite{Vermaseren:1998uu,Blumlein:1998if,Ablinger:2013cf,Ablinger:2013jta,Ablinger:2014rba} have reached for single-scale quantities -- such as DIS \cite{Moch:1999eb}, Drell-Yan \cite{Blumlein:2005im}, splitting functions \cite{Vogt:2004,Moch:2004}, and Higgs-production \cite{Ball:2013bra} -- and the status of genuine double Mellin transforms.

In the following, we want to illustrate how the results in terms of SVPs may help in establishing double Mellin transforms for SIDIS at NNLO.
As an example, we look at the coefficient function $\mathcal{C}_{gq}^\text{T}$.
This contains several SVPs, here we will discuss the part which depends on $\svp_2\!\left(\frac{1-x}{1-z}\right)$,
concretely
	\begin{align}
		\!\!\left.\tilde{\mathcal{C}}_{gq}^\text{T}\right|_{\svp_2}\!\!\!\!(N,M)=\!\int_0^1\!\dx x\,x^{N-1} \!\int_0^1\!\dx z\,z^{M-1} f(x,z) \svp_2\!\left(\frac{1-x}{1-z}\right)
		\label{eq: Mellin-intergral SVP}
	\end{align}
with
	\begin{align}
		f(x,z)=&\NC^2\left[x z+\frac{x}{2 z}-\frac{z}{2 (1-x)}-\frac{1}{(1-x) z}-x\right.\nonumber\\
		 &\qquad+\left.\frac{1}{1-x}+\frac{1}{2 z}-1\right]
		+\frac{1}{1-x}-\frac{z}{2 (1-x)}\nonumber\\
		-&\frac{1}{\NC^2}\left[x z+\frac{x}{2 z}-\frac{z}{1-x}-\frac{1}{(1-x) z}-x\right.\nonumber\\
		 &\qquad+\left.
		\frac{2}{1-x}+\frac{1}{2 z}-1\right].
	\end{align}
	
To calculate eq.\,\eqref{eq: Mellin-intergral SVP}, it is sensible to convert the SVP into Goncharov polylogarithms $G(a_1,\dots,a_n;t)$ \cite{Goncharov:2001}, allowing for the subsequent utilization of \texttt{PolyLogTools} \citep{Duhr:2019}.
For this, it is most convenient to map the argument of the SVP onto the unit interval using functional equations (see Appendix C of \cite{Haug:2022}) and then use the representation
\begin{align}
\svp_2(t)=-G(0,1;t)+\frac{1}{2}G(0;t)G(1;t)\quad\text{for}\quad 0<t<1\,.
\label{eq: SVP Goncharov representation}
\end{align}

In the present case, there is the slight subtlety that the prefactor $f(x,z)$ contains a singularity $1/(1-x)$ that is regularized by the zero of the SVP.
The simplest way to deal with this issue is to split the $x$ integral into the region $z<x<1$, which contains the singularity and where the SVP argument lies between $0$ and $1$, and the region $0<x<z$, which is free of singularities.
In the latter we are thus free to use functional equations on the SVP and split the resulting terms, explicitly we apply the inverted Landen identity \cite{Haug:2022}
\begin{align}
\svp_2\left(\frac{1-x}{1-z}\right)=\svp_2\!\left(1-\frac{1-z}{1-x}\right)+\zeta_2.
\end{align}
Substituting the arguments of the SVP as the new integration variable, we obtain the representation
\begin{widetext}
	\begin{align}
		\left.\tilde{\mathcal{C}}_{gq}^\text{T}\right|_{\svp_2}\!\!\!\!(N,M)=\!\int_0^1\!\dx z\,z^{M-1}&\left\lbrace
		\int_0^z\dx t\, \frac{1-z}{(1-t)^2} \left(\frac{z-t}{1-t}\right)^{N-1} f\!\left(\frac{z-t}{1-t},z\right) \left[\svp_2(t)+\zeta_2\right]
		\right.\nonumber\\
		&\left.
		+(1-z)\int_0^1\dx t\,(1-t-zt)^{N-1} f(1-t-zt,z)\svp_2(t)
		\hphantom{\frac{}{}}\right\rbrace\,.
		\label{eq: Double Mellin trafo example}
	\end{align}
Note that, so far, this representation is analytic in $N$ and $M$.
From here, plugging in the representation \eqref{eq: SVP Goncharov representation}, we can directly calculate the Mellin moments for fixed integer values of $N$ and $M$ with \texttt{PolyLogTools},
	\begin{align}
	\left.\tilde{\mathcal{C}}_{gq}^\text{T}\right|_{\svp_2}\!\!\!\!(1,2)
	&=\NC^2 \left(-\zeta_3-\frac{1}{32}-\frac{31 \pi ^2}{240}\right)+\frac{1}{\NC^2} \left(\frac{\zeta_3}{2}+\frac{1}{96}+\frac{13 \pi ^2}{720}\right)+\frac{\zeta_3}{2}+\frac{\pi ^2}{9}+\frac{1}{48}\,,\nonumber\\ \left.\tilde{\mathcal{C}}_{gq}^\text{T}\right|_{\svp_2}\!\!\!\!(1,3)
	&=\NC^2 \left(-\frac{7 \zeta_3}{16}-\frac{719}{13824}-\frac{611 \pi ^2}{5760}\right)+\frac{1}{\NC^2} \left(\frac{\zeta_3}{8}+\frac{97}{6912}+\frac{53 \pi ^2}{2880}\right)+\frac{5 \zeta_3}{16}+\frac{101 \pi ^2}{1152}+\frac{175}{4608}\,,\nonumber\\ \left.\tilde{\mathcal{C}}_{gq}^\text{T}\right|_{\svp_2}\!\!\!\!(2,2)
	&=\NC^2 \left(-\zeta_3-\frac{1}{32}-\frac{59 \pi ^2}{2160}\right)+\frac{1}{\NC^2} \left(\frac{\zeta_3}{2}+\frac{5}{192}-\frac{23 \pi ^2}{1080}\right)+\frac{\zeta_3}{2}+\frac{7 \pi ^2}{144}+\frac{1}{192}\,,\nonumber\\ \left.\tilde{\mathcal{C}}_{gq}^\text{T}\right|_{\svp_2}\!\!\!\!(1,4)
	&=\NC^2 \left(-\frac{11 \zeta_3}{40}-\frac{2077}{34560}-\frac{2953 \pi ^2}{33600}\right)+\frac{1}{\NC^2} \left(\frac{\zeta_3}{20}+\frac{17}{1080}+\frac{31 \pi ^2}{2100}\right)+\frac{9 \zeta_3}{40}+\frac{117 \pi ^2}{1600}+\frac{511}{11520}\,,\nonumber\\ \left.\tilde{\mathcal{C}}_{gq}^\text{T}\right|_{\svp_2}\!\!\!\!(2,3)
	&=\NC^2 \left(-\frac{7 \zeta_3}{16}-\frac{271}{13824}-\frac{5543 \pi ^2}{120960}\right)+\frac{1}{\NC^2} \left(\frac{\zeta_3}{8}+\frac{47}{6912}+\frac{31 \pi ^2}{60480}\right)+\frac{5 \zeta_3}{16}+\frac{29 \pi ^2}{640}+\frac{59}{4608}\,,\nonumber\\ \left.\tilde{\mathcal{C}}_{gq}^\text{T}\right|_{\svp_2}\!\!\!\!(3,2)
	&=\NC^2 \left(-\zeta_3-\frac{55}{864}+\frac{\pi ^2}{70}\right)+\frac{1}{\NC^2} \left(\frac{\zeta_3}{2}+\frac{23}{432}-\frac{23 \pi ^2}{630}\right)+\frac{\zeta_3}{2}+\frac{\pi ^2}{45}+\frac{1}{96}\,,
	\nonumber\\
	&\vdots
	\nonumber\\
	\nonumber\\
	&\phantom{vdots}\nonumber\\
	&\vdots
	\nonumber\\
	\left.\tilde{\mathcal{C}}_{gq}^\text{T}\right|_{\svp_2}\!\!\!\!(25,25)
	&=\NC^2 \left(-\frac{163 \zeta_3}{5200}-\frac{191603034322639217188141108951}{82191180313902788098274304000000}-\frac{343704078376116047397239 \pi ^2}{145035282798712645859520000}\right)
	\nonumber\\&+\frac{1}{\NC^2} \left(\frac{\zeta_3}{5200}+\frac{2626501038131451153343876273}{82191180313902788098274304000000}-\frac{1162009044070230027019 \pi ^2}{145035282798712645859520000}\right)
	\nonumber\\&+\frac{81 \zeta_3}{2600}+\frac{163754077597429381493 \pi ^2}{68867655649911037920000}+\frac{68618930023423299213797107}{29844292052978499672576000000}\,.
	\end{align}
\end{widetext}

Similar strategies may be applied to the other terms containing SVPs in all coefficient functions, however a detailed study is beyond the scope of this work.
Here, it might be helpful to note that all appearing SVPs with arguments ``quadratic" in $x$ or $z$ can be removed using Abel's five-term relation for the single-valued dilogarithm, explicitly

\begin{widetext}
\begin{align}
\mathcal{L}_2\!\left(\frac{x(1-z)}{z(1-x)}\right)&=
-\mathcal{L}_2(x)+\mathcal{L}_2(z)+\mathcal{L}_2\!\left(\frac{x}{z}\right)
-\mathcal{L}_2\!\left(\frac{1-x}{1-z}\right)+\zeta_2\,,\\
\mathcal{L}_2\!\left(\frac{x(1-x)}{z(1-z)}\right)&=
\mathcal{L}_2\!\left(\frac{1-x}{z}\right)-\mathcal{L}_2\!\left(\frac{1-z}{x}\right)
+\mathcal{L}_2\!\left(\frac{1-x}{1-z}\right)
+\mathcal{L}_2\!\left(\frac{x}{z}\right)-\zeta_2\,,\\
\mathcal{L}_2\!\left(\frac{x z}{(1-x)(1-z)}\right)&=
-\mathcal{L}_2(x)-\mathcal{L}_2(z)-\mathcal{L}_2\!\left(\frac{1-x}{z}\right)
-\mathcal{L}_2\!\left(\frac{1-z}{x}\right)+4\zeta_2\,.
\end{align}
\end{widetext}

Of course, the SVP terms are only a small part of the complete coefficient functions and there are other terms posing certain challenges in the integration, notably dilogarithms with square-root arguments. 
Nevertheless, the representation in terms of SVPs allows for a very transparent treatment of the terms with case distinctions.
This is also aided by the clean functional equations satisfied by $\svp_2$, which in the presented case permitted for a ``safe'' mapping on Goncharov polylogarithms away from branching points.

If feasible, going beyond fixed moments and reconstructing the double Mellin transform in the complex plane would be the ideal scenario going forward.
A promising strategy here might be to construct a sufficiently general ansatz, presumably in the space of nested sums, and fit (rational) coefficients to the pre-computed grid of integer moments. 
If successful, the analytic continuation is reduced to the known continuation of the ansatz functions. 
\bibliography{SIDIS_with_SVPs.bib}

\end{document}